# Epitaxial thin films of pyrochlore iridate Bi$_{2+x}$Ir$_{2-y}$O$_{7-\delta}$: structure, defects and transport properties

W. C. Yang[1,+], Y. T. Xie[2,+], W. K. Zhu[1,+], K. Park[2], A. P. Chen[3], Y. Losovyj[4], Z. Li[1,3], H. M. Liu[1], M. Starr[1], J. A. Acosta[1], C. G. Tao[2], N. Li[3], Q. X. Jia[3,5], J. J. Heremans[2], and S. X. Zhang[1,*]

[1]Department of Physics, Indiana University, Bloomington, Indiana 47405, USA
[2]Department of Physics, Virginia Tech, Blacksburg, Virginia 24061, USA
[3]Center for Integrated Nanotechnologies, Los Alamos National Laboratory, Los Alamos, 87545 USA
[4]Department of Chemistry, Indiana University, Bloomington, Indiana 47405, USA
[5]Department of Materials Design and Innovation, University at Buffalo, The State University of New York, Buffalo, NY 14260, USA
* sxzhang@indiana.edu
[+] these authors contributed equally to this work

## Abstract

While pyrochlore iridate thin films are theoretically predicted to possess a variety of emergent topological properties, experimental verification of these predictions can be obstructed by the challenge in thin film growth. Here we report on the pulsed laser deposition and characterization of thin films of a representative pyrochlore compound Bi$_2$Ir$_2$O$_7$. The films were epitaxially grown on yttria-stabilized zirconia substrates and have lattice constants that are a few percent larger than that of the bulk single crystals. The film composition shows a strong dependence on the oxygen partial pressure. Density-functional-theory calculations indicate the existence of Bi$_{Ir}$ antisite defects, qualitatively consistent with the high Bi: Ir ratio found in the films. Both Ir and Bi have oxidation states that are lower than their nominal values, suggesting the existence of oxygen deficiency. The iridate thin films show a variety of intriguing transport characteristics, including multiple charge carriers, logarithmic dependence of resistance on temperature, antilocalization corrections to conductance due to spin-orbit interactions, and linear positive magnetoresistance.

## Introduction

Iridates have recently emerged as a fertile ground for novel topological electronic states that arise from the interplay of electron interactions and spin-orbit coupling.[1-9] In particular, the pyrochlore compounds with a chemical formula of $A_2Ir_2O_7$ (A-227, where A=Bi, Y or rare-earth element) are predicted to host a variety of topological phases, including topological Mott insulators,[3,4] Weyl semimetals,[6] topological crystalline insulators,[10] and Weyl-Mott insulators.[11] The pyrochlore iridate compounds have a cubic crystal structure ($Fm\bar{3}m$), in which the $A^{3+}$ and $Ir^{4+}$ cations form inter-penetrating networks of corner-sharing tetrahedra. The energy scale of electron interaction relative to electron hopping (i.e. U/t) increases monotonically as the A-site ionic radius decreases;[12] and the electrical transport properties can be divided roughly into three categories:[8,13] 1) compounds with large $A^{3+}$ (e.g. Bi-227 and Pr-227) show a metallic behavior down to the lowest measured temperature $T$;[5,13-16] 2) those with intermediate $A^{3+}$ (e.g. Eu-227 and Nd-227) display a metal-to-insulator transition,[13,16-19] which is often accompanied by a paramagnetic to an antiferromagnetic phase transition; and 3) substances with small $A^{3+}$ (e.g. Lu-227 and Yb-227) exhibit an insulating-like behavior[13,15,20] throughout the entire region of $T$.

While significant experimental progress has recently been made in bulk A-227 compounds,[5,17-38] including the demonstration of all-in/all-out non-collinear magnetic order,[24,38] observation of giant magnetoresistance associated with metallic domain walls[34-36] and indication of Weyl semimetal phases,[18,37] there are only a few studies on thin film samples,[39-43] which is largely due to the great challenges in obtaining high quality films. Theoretical studies on pyrocholore iridate thin films have suggested a variety of emergent topological properties, including the quantized anomalous Hall conductance[44] and the correlated Chern insulator[45] that are otherwise not accessible in bulk samples. While these topological properties have yet been realized experimentally, remarkable transport phenomena have been observed in thin films, including a linear magnetoresistance (MR) up to 35 Tesla in Bi-227[39] and a cooling field-dependent, assymetric MR in Eu-227.[40] Moreover, thin film structures provide an excellent platform to engineer physical properties by means of, for example, electric gating and elastic strain, offering great potentials for device applications.

In this paper, we report on systematic measurements of the structural, compositional, and electrical transport properties of a prototypical pyrochlore iridate, namely Bi-227 thin films. The films were grown epitaxially along the [111] direction on the yttrium-stabilized zirconia (YSZ) (111) substrates by pulsed laser deposition (PLD). The out-of-plane lattice parameters of thin films were found to be a few percent larger than that of the bulk samples. Compositional characterizations, density-functional-theory (DFT) calculations, along with X-ray photoelectron spectroscopy (XPS) studies suggest the existence of $Bi_{Ir}$ antisite defects and oxygen vacancies. With varied defect densities, thin films show a broad spectrum of electronic transport characteristics, including multiple charge carriers, logarithmic dependence of resistance on *T*, antilocalization, and linear positive magnetoresistance.

## Results and Discussions

### PLD growth, structural and compositional characterizations

Thin films were grown by PLD using two targets that were prepared via a solid state reaction method with $IrO_2$ and $Bi_2O_3$ as source materials at Ir/Bi ratios of 1 and 3 (labeled as Ir/Bi@1 target and Ir/Bi@3 target, respectively). X-ray diffraction (XRD) measurements suggest that thin films are epitaxially grown on the YSZ substrates along the [111] direction of a cubic phase. Figure 1(a) shows a local $2\theta - \omega$ scan taken on a representative film grown at oxygen partial pressure $P_{O2}$ = 50 mTorr using the Ir/Bi@1 target. A well-defined oscillation was observed near the thin film (222) peak, indicating a high crystalline quality and smooth film surface. The film thickness was determined to be ~ 26 nm based on the X-ray reflectivity. The out-of-plane *d*-spacing for the (222) film peak ($d_{222}$) is ~ 3.07 Å, which is ~ 3% larger than that of bulk Bi-227, i.e., $d_{222}$ ~ 2.98 Å.[14,29] The in-plane $\varphi$-scan shows a clear three-fold symmetry [Fig. 1 (b)], confirming that the film has a cubic crystalline structure and is epitaxially grown on the substrate. The epitaxial relationship between the thin film and the substrate was determined to be (222)Bi-227∥(111)YSZ and [02$\bar{2}$]Bi-227∥[01$\bar{1}$]YSZ based on the $\varphi$-scan and the cross-sectional transmission electron microscopy (TEM) measurements [Fig.1 (c) (d) and (e)].

We note that elongation (contraction) in the out-of-plane *d*-spacing was also observed in a layered iridate $Sr_2IrO_4$ and its origin was demonstrated to be in-plane compression (tension).[46,47] While a compressive in-plane strain may exist in the Bi-227 film due to its lattice mismatch with the substrate ($a_{bulk\ Bi-227}$ ~10.32 Å versus $2a_{YSZ}$ ~ 10.25 Å), the elongation in out-of-plane *d*-spacing here cannot be solely attributed to the in-plane compression. Indeed, a 1.5% in-plane compression is required for 3% out-of-plane elongation in order to maintain the unit cell volume constant. However, the lattice mismatch between the film (assuming bulk lattice constant) and the substrate is only 0.6%, smaller than the required compression. To understand the origin of the large lattice elongation, we further carried out film growth at different $P_{O2}$ using two different targets since oxygen content and cation stoichiometry can be strongly correlated with the out-of-plane lattice parameters in complex oxide thin films.[48-50] As shown in Fig. 2 (a), the spacing $d_{222}$ (and hence the lattice constant $a = 2\sqrt{3}d_{222}$ of the cubic unit cell) decreases with the decrease of $P_{O2}$. At the same $P_{O2}$, the film grown using the Ir/Bi@3 target has a smaller lattice constant than that grown using the Ir/Bi@1 target. The minimum $d_{222}$ (or $a$) obtained in these films is still ~ 1.3% larger than that of the bulk sample. To gain further insight, we characterized the chemical compositions of the films using two complementary techniques, i.e. XPS and energy-dispersion X-ray spectroscopy (EDX) which show consistent dependence of Ir/Bi ratio on $P_{O2}$ [Fig. 2 (b)]. The ratio increases from ~ 0.4 to ~ 1 when $P_{O2}$ is decreased from 50 mTorr to 10 mTorr. The low Ir/Bi ratio at high $P_{O2}$ should be attributed to the high vapor pressure of $IrO_3$. We also note that a Pt/Bi ratio of ~0.62 (or Bi/Pt ~1.62) was determined in another pyrochlore film $Bi_2Pt_2O_7$;[51] these results suggest that the pyrochlore phase is stable over a wide range of compositions. When $P_{O2}$ is 30 mTorr or lower, a small Ir (111) peak is observed in the XRD pattern (see Supplementary Fig. S1), indicating the formation of Ir metal impurity. The film grown at $P_{O2} = 1$ mTorr is dominated by Ir metal and the Ir/Bi ratio is found to be > 10 (see Supplementary Fig. S1 and S2 (c)). The high Ir/Bi ratio obtained in low $P_{O2}$ may be attributed to the high volatility of Bi metal and stability of Ir. At the same $P_{O2}$, the films deposited from the Ir/Bi@3 target have slightly higher Ir/Bi ratios than the ones from the Ir/Bi@1 target. Nevertheless, the dependence of film composition on the target is much weaker than on $P_{O2}$.

**Density Functional Theory Calculations**

The low Ir/Bi ratio in the films grown at high $P_{O2}$ indicates the possible existence of antisite ($Bi_{Ir}$) and/or Bi interstitial ($Bi_{int}$) defects which can result in a larger lattice constant than the stoichiometric compound (the ionic radius of $Bi^{3+}$ is larger than $Ir^{4+}$). To verify this possibility, we computed the formation energies of the above two types of defects via DFT calculations, in which a range of chemical potential differences of Bi, Ir and O are considered [Table 1]. When the chemical potential of elemental Bi, Ir, or O is the same as that of bulk Bi, bulk Ir, or $O_2$ gas, respectively, i.e. the chemical potential difference is zero, the Bi, Ir, or O is said to be in an " abundant" condition. A decrease of chemical potential difference corresponds to the decrease of element content; in other words, the lower the chemical potential difference is, the "more deficient" the element is. Our calculations show that the $Bi_{int}$ defects have a consistently high formation energy (4.86 ~6.50 eV), indicating a low possibility of this defect in the films. In contrast, the formation energy of the antisite $Bi_{Ir}$ defect varies from negative to positive, depending mainly on the oxygen content or the chemical potential difference between the element O and $O_2$ gas. When oxygen is abundant (cases # 1 and 2) or slightly deficient (cases # 3 and 4), the formation energy is found to be negative. Comparing case # 1 with # 2 (or # 3 with # 4) suggests that the more deficient the Ir is, the more stable the $Bi_{Ir}$ defect is; furthermore, comparison between cases # 1 and 3 (or between # 2 and 4) indicates that the more abundant the oxygen is, the more stable the defect is. Particularly, in case # 1 when oxygen is abundant and Ir is deficient, the formation energy reaches a minimum value of -7.47 eV, suggesting that the antisite defect is very stable in this condition. The calculated result is qualitatively consistent with the experimental result in Fig. 2 (a), i.e. the Ir/Bi ratio in the film increases as $P_{O2}$ decreases. When oxygen is very deficient but Bi and Ir are abundant (case # 5), the antisite defect is unlikely to form as indicated by a positive formation energy of 4.04 eV, consistent with the high Ir/Bi (>10) observed at the lowest $P_{O2}$ [see Supplementary Fig. S2 (c)].

**X-ray photoelectron spectroscopy studies of oxidation states**

The films grown at $P_{O2}$ = 15 ~ 50 mTorr were found to be slightly oxygen-deficient, as suggested by the XPS studies. The XPS measurement was taken on three representative films, i.e. f1 ($P_{O2}$= 50 mTorr, Ir/Bi@1 target), f2 ($P_{O2}$= 50 mTorr, Ir/Bi@3 target), and f3 ($P_{O2}$= 15 mTorr, Ir/Bi@3 target), along with two control samples, i.e. bulk $IrO_2$ and $Bi_2O_3$. The Ir 4f spectrum and Bi 4f spectrum for films f1 and f3 are presented in Fig. 3, and the same for f2 are provided in the supplementary Fig. S3 online. A qualitative comparison between the spectra of the films and that of the bulk $IrO_2$ (and $Bi_2O_3$) (see supplementary Fig. S4) suggests that both Ir (and Bi) have a component with an oxidation state lower than the nominal +4 (and +3). We fitted the spectra using two components in CasaXPS software to obtain more quantitative information. The symmetric component is described by the Gaussian-Lorentz profile GL(m), while the asymmetric lineshape which is used to capture the many-body, screening effects[52] is described by the convolution of Gelius profile A(a,b,n) and Gaussian-Lorentz profile GL(m), in which the assymetry is determined by the parameters a and b. More detailed information about the fitting process is provided in the supplemental information. The fitting spectra at the optimal a and b values are shown in Figure 3. The obtained binding energies for Ir $4f_{5/2}$ peaks are: 65.8~66.1 eV (component 1) and 64.4~64.5 eV (component 2); the binding energies for Ir $4f_{7/2}$ peaks are: 62.8~63.1 eV (component 1); and 61.4~61.5 eV (component 2). The existence of two components may be attributed to the apperance of two final states in the photoemission process,[53,54] the coexistence of the core lines and plasmon satellite,[55,56] and a mixed oxidation state.[57] In either case, however, the fact that the entire measured spectra are shifted to lower binding energies in comparison to $IrO_2$ ($Ir^{4+}$) suggests the formation of oxygen vacancies in the films. Similarly, the Bi spectrum can be fitted using two components as well [Fig. 3 (b)]: binding energies of 163.8~164.1 eV (component 1) and 163.0~163.2 eV (component 2) for the $4f_{5/2}$ peak; and 158.5~158.8 eV (component 1) and 157.7~157.9 eV (component 2) for the $4f_{7/2}$ peak. Again the binding energies being lower than for $Bi^{3+}$ indicates the presence of oxygen deficiency.

The existence of $Bi_{Ir}$ defects and oxygen vacancies in our pyrochlore thin films is not too surprising. First, cation antisite defects have already been observed in other pyrochlore compounds such as $Gd_2Zr_2O_7$.[58] Second, the structure of pyrochlore $Bi_2Ir_2O_7$ is remarkably similar to that of the cubic δ–Bi$_2$O$_3$ [59] which can be viewed as Bi$_2$Bi$_2$O$_6$. Figure 4 shows a unit cell of the cubic δ–Bi$_2$O$_3$, a 1/8 unit cell of Bi$_2$Ir$_2$O$_7$ with and without anitsite defects and oxygen vacancies. In comparison to Bi$_2$Ir$_2$O$_7$, the cubic δ–Bi$_2$O$_3$ has one less oxygen atom and all Ir sites are occupied by Bi. As a result, a signicant amount of Bi$_{Ir}$ defects and oxygen vacancies can exist in Bi$_2$Ir$_2$O$_7$ while maintaining its cubic crystal structure.

**Transport measurements and discusion**

With the existence of antisite defects and oxygen vacancies, the Bi$_{2+x}$Ir$_{2-y}$O$_{7-\delta}$ thin films show revealing electronic transport properties. The dependence on $T$ of the sheet resistance (2D resistivity) $R_\square$ at magnetic field $B = 0$ and the low-$T$ magnetoresistance (MR, $R_\square$ vs $B$) and the Hall resistance $R_H$ vs $B$ were obtained on Hall bars (inset in Figure 5 and *cfr*. Methods) prepared by photolithography and dry etching on the three representative films, f1, f2, and f3. Values for $R_\square$ at $B = 0$ and $T = 0.39$ K are, 483 Ω/□ (f1), 1334 Ω/□ (f2), and 1013 Ω/□ (f3). As observed in Figure 5 at $T = 0.39$ K, $R_H$ shows a non-linear dependence on $B$ for all films, suggesting the existence of multiple types of charge carriers. In particular, $R_H(B)$ of f1 can be fitted to a two-carrier model by including both electrons and holes, while f3 is found to host two types of holes with different carrier mobilities. The sheet charge carrier densities ($n_e$ for electrons, $n_h$ for holes) and mobilities ($\mu_e$ for electrons, $\mu_h$ for holes) determined from the fittings at $T = 0.39$ K are: $n_e = 8.0 \times 10^{20} m^{-2}, \mu_e = 1.5 \times 10^{-5} m^2 V^{-1} s^{-1}$ and $n_h = 2.0 \times 10^{12} m^{-2}, \mu_h = 0.38 m^2 V^{-1} s^{-1}$ for film f1; $n_h = 6.4 \times 10^{20} m^{-2}, \mu_h = 1.0 \times 10^{-5} m^2 V^{-1} s^{-1}$ and $n_h = 3.0 \times 10^{12} m^{-2}, \mu_h = 0.23 m^2 V^{-1} s^{-1}$ for film f3. $R_H(B)$ for f2 cannot be fitted to a two-carrier model, yet the use of more than two types of carriers will lead to uncertainty in the fitting values due to the proliferation of fitting parameters. While the complicated $R_H(B)$ for f2 indicates the presence of multiple carriers, its slope indicates that the dominant charge carriers are holes, with average $n_h = 1.4 \times 10^{20} m^{-2}$.

The clear experimental evidence for the presence of multiple charge carriers in films f1, f2 and f3 is qualitatively consistent with the existence of multiple bands in the electronic structures.[60] The multiple bands will impart different effective masses, qualitatively consistent also with the experimental evidence for different mobilities. While oxygen vacancies are expected to function as *n*-type dopants and $Bi_{Ir}$ antisite defects as *p*-type dopants, it is less certain that the observed carrier densities should be associated with doping via these different types of defects since thermal ionization of defect levels may not be effective at $T = 0.39$ K. At the high defect density present in the films, *e.g.* where 1/3 of Ir is replaced by Bi, the defects can be expected to impact the electronic band structure instead of solely change the Fermi level via doping, not captured in DFT calculations based on assumed stoichiometry. Association between observed transport properties and specific defects therefore requires future theoretical and experimental studies of band structure in the presence of high defect densities.

All three films show metallic behavior in $R_\square$ *vs* $T$ at high $T$, as Figure 6 shows. At intermediate $T$, the films show a logarithmic increase of $R_\square$ with decreasing $T$ such that $1/R_\square \sim \ln(T/T_0)$ (Figure 6). The transition from metallic to logarithmic insulating behavior occurs for f1 at 10 K, for f2 at 50 K and for f3 at 55 K. It is apparent that among the three films, f1 maintains its metallic character to the lowest $T$, compatible with a higher crystalline quality. Indeed, XRD for f1 shows the sharpest features (see Supplementary Fig. S1), a sign of the best crystalline quality among the films. The logarithmic dependence of the sheet conductance (2D conductivity) $G_\square = 1/R_\square$ on $T$, depicted in Figure 7, can at $B = 0$ be expressed as:

$$G_\square = G_{\square 0} + \alpha(e^2/(2\pi^2\hbar)) \ln(T/T_0) \qquad (1)$$

where $G_{\square 0}$ denotes a *T*-independent part, $T_0$ is a normalization constant, and $\alpha$ denotes a prefactor. Figure 7 depicts $G_\square$ (*B*=0) *vs* $T$ on a semi-logarithmic graph, with fits to Equation (1) yielding $\alpha =$ 0.67 for f1, $\alpha = 0.71$ for f2 and $\alpha = 0.87$ for f3. The values for $\alpha$ are typical for thin films in general, where $\alpha$ of order unity is most often encountered.[61] Two mechanisms can lead to a logarithmic dependence but in all present films have opposite effects on $G_\square$: antilocalization which will tend to increase $G_\square$ with decreasing $T$ at $B = 0$, and electron-electron interactions which will tend to decrease $G_\square$ with decreasing $T$. Antilocalization results from the destructive interference of partial waves on time-reversed paths returning to the origin of the paths, occurring in the presence of spin-orbit interaction when the mobility mean-free-path is shorter than the quantum phase coherence length.[61-66] Antilocalization is accompanied by a characteristic positive MR, observed in the films.[61-66] Figure 8 shows the MR obtained at $T = 0.39$ K for the films. A pronounced low-$B$ positive MR characteristic of antilocalization is observed for all three films, transitioning to a linear positive MR at high $B$ (discussed below). The existence of antilocalization confirms the presence of spin-orbit interaction in the films, and accounts for the observation of relatively low $\alpha < 1$. As a quantum coherence effect, the antilocalization correction to the classical sheet conductance strengthens with decreasing $T$, which increases $G_\square$ with decreasing $T$. On the other hand, electron-electron interactions can directly decrease $G_\square$ ($B=0$) with decreasing $T$ via the Aronov-Altshuler mechanism,[67] where the interaction leads to an effective suppression of the density-of-states at the Fermi level in diffusive transport. The Aronov-Altshuler mechanism results in a logarithmic dependence on $T$ as expressed in Equation (1) with $\alpha$ ranging from 0.25 (strong screening of electron-electron interactions) to 1 (no screening).[61,66] The Aronov-Altshuler mechanism is also accompanied by a weak positive MR.[61] In the present films the dependence $G_\square \sim \ln(T/T_0)$ down to $T \approx 1$ K with prefactor $\alpha < 1$, and the upturn in $G_\square$ for $T < 1$ K are attributed to the competing effects of antilocalization and the Aronov-Altshuler electron-electron interaction mechanism. The observation of positive MR at low $B$ (Figure 8) is further consistent with the presence of both antilocalization and the Aronov-Altshuler mechanism. A detailed analysis of the exact contribution of each effect is outside the scope of this work. The linear positive MR at higher $B > 1.5$ T in Figure 8 is similar to the MR observed in previous work on thin film Bi-227,[39] although no hysterisis was observed in our thin films. We ascribe the linear positive MR at high $B$ to the underlying electronic structure featuring a linear dispersion. Although its exact origins are still debated, a positive linear MR has indeed consistently appeared in the context of materials with a linear dispersion or quasi-relativistic dispersion (Dirac materials, topological insulators, Bi, InSb).[68]

## Conclusions

In conclusion, we have achieved epitaxial growth of pyrochlore iridate thin films on yttria-stabilized zirconia substrate via pulsed laser deposition. The lattice constants of the films are a few percent larger than that of the bulk single crystal, and the film composition shows a strong dependence on the oxygen partial pressure $P_{O2}$. DFT calculations indicate the existence of $Bi_{Ir}$, which is qualitatively consistent with the large lattice constant and low Ir/Bi ratio found in the films grown at a relatively high $P_{O2}$. Both Ir and Bi have oxidation states that are lower than their nominal values, suggesting the existence of oxygen deficiency. With antisite defects and oxygen vacancies, the $Bi_{2+x}Ir_{2-y}O_{7-\delta}$ thin films show a variety of intriguing electronic transport properties, including multi-carrier transport, a logarithmic dependence of conductance on $T$, an antilocalization quantum correction to conductance due to spin-orbit interaction, and linear positive magnetoresistance.

## Methods

### Experimental details

Thin films of $Bi_{2+x}Ir_{2-Y}O_{7-\delta}$ were grown on yttria-stabilized zirconia (YSZ) (111) substrates by PLD. Two ceramic targets were prepared via a solid state reaction method using $IrO_2$ and $Bi_2O_3$ as source materials at Ir/Bi ratios of 1 and 3 (labeled as Ir/Bi@1 target and Ir/Bi@3 target, respectively). The repetition rate of the KrF excimer laser ($\lambda = 248$ nm) was 1 Hz and the nominal energy density was ~ 3.33 J/cm$^2$. The substrate heater temperature was set to 750 °C (actual substrate temperature ~550 °C), and $P_{O2}$ was varied from 1 to 50 mTorr. Films grown at three representative conditions labeled as f1 ($P_{O2}$= 50 mTorr, Ir/Bi@1 target); f2 ($P_{O2}$ = 50 mTorr, Ir/Bi@3 target); and f3 ($P_{O2}$ = 15 mTorr, Ir/Bi@3 target), were the focus of this work. After deposition, the PLD chamber was filled with oxygen gas up to about atmosphere pressure, in which thin films were cooled down. X-ray diffraction (Panalytical X'Pert PRO MRD) $2\theta$-$\omega$ (Triple axis mode) and φ-scans (Rocking curve mode) were carried out in Los Alamos National Laboratory to obtain information on the orientation, lattice parameters and epitaxial quality of the thin films. Further XRD measurements were also carried out in Indiana University using a standard PANalytical instrument with a Chi-Phi-Z sample stage (Cu K$_\alpha$). Cross-section specimens for TEM were prepared by mechanical polishing of the film to a final thickness of ~ 60 μm with a diamond lapping film, followed by a thinning process using a Gatan Precision Ion Polishing System Model 691. TEM characterization was conducted in an FEI Tecnai F30 transmission electron microscope. Energy-dispersion X-ray spectroscopy (EDX) measurement was carried out in a scanning electron microscope (SEM, Quanta FEI). X-ray photoelectron spectroscopy (XPS) data were obtained on a PHI VersaProbe II Scanning X-ray Microprobe system. All XPS spectra were calibrated using the carbon 1s peak at 284.8 eV. Magnetotransport measurements were performed on microfabricated Hall bars (*cfr*. inset in Figure 5), obtained by photolithography followed by reactive ion etching in BCl3. The Hall elements were L-shaped to enable characterization of anisotropic transport properties (anisotropy was not observed). The active region of the Hall elements had a length-to-width ratio of 2 (160 µm length, 80 µm width), sufficiently high to allow observation of a clear Hall signal despite the low values for $R_H(B)$. Ohmic contacts were photolithographically fabricated as unannealed pads of 5 nm Cr / 40 nm Au. Magnetotransport was measured over 390 mK < $T$ < 270 K in a sample-in-liquid $^3$He system. Excitation currents varied between 5 µA (for $R_\square$) to 20 µA (for $R_H$), sufficiently low to avoid carrier heating. Zero-field resistances were measured during cool-down from 270 K to 390 mK with magnetic field $B$ = 0 T (ZFC). Magnetoresistances were then measured at 390 mK over -9 T< $B$ < 9 T. Samples were subsequently brought to $T$ = 30 K and $B$ = 4.0 T and cooled to $T$ = 390 mK at $B$ = 4.0 T (FC). Magnetoresistances were then remeasured at 390 mK over -9 T< $B$ < 9 T. Differences in data between ZFC and FC conditions were not apparent and hence magnetotransport data obtained under ZFC conditions only are shown.

**DFT clculation details**

Density-functional theory calculations were carried out on the pyrochlore structure with and without defects by using DFT code, VASP.[69,70] The generalized gradient approximation (GGA)[71] was used for exchange-correlation functional and projector-augmented wave (PAW) pseudopotentials.[72] Spin-orbit coupling was included self-consistently within the DFT calculation. For the perfect Bi-227 pyrochlore structure, we considered a face-centered cubic (fcc) primitive unit cell of 22 atoms with the experimental lattice constant of 5.155 Å[14] and relaxed the geometry until the residual forces became less than 0.01 eV/Å. The energy cutoff of 400 eV and 9×9×1 $k$-point mesh were used for the relaxation and the self-consistent run of the optimized geometry.

For the structure with defects, we considered two different types of defects, i.e. Bi: Ir antisite ($Bi_{Ir}$) and Bi interstitial ($Bi_{int}$) that could result in the high Bi to Ir ratio found in the experiment. In each defect type, we simulated an 88-atom supercell with one defect when the structure is electrically neutral (see Supplementary Fig. S6 online). Possible oxygen vacancies were not introduced in the structure. In the structure with defects, the geometry was relaxed with 5×5×1 $k$-point mesh and an energy cutoff of 400 eV until the residual forces were less than 0.01 eV/Å. The formation energy of a Bi antisite defect is $\Delta E_f = E_{anti} - E_0 - (\mu_{Bi}^{bulk} + \Delta\mu_{Bi}) + (\mu_{Ir}^{bulk} + \Delta\mu_{Ir})$, whereas the formation energy of an interstitial Bi defect is $\Delta E_f = E_{int} - E_0 - (\mu_{Bi}^{bulk} + \Delta\mu_{Bi})$, where $E_{anti}$, $E_{int}$, and $E_0$ are total energies of the structure with an antisite defect, with an interstitial defect, and without any defects, respectively.[73-75] Here $\mu_{Bi}^{bulk}$ and $\mu_{Ir}^{bulk}$ are the chemical potential of bulk Bi and Ir, while $\Delta\mu_{Bi}$ and $\Delta\mu_{Ir}$ are the chemical potential differences from their bulk values. The chemical potentials of bulk Bi and Ir were calculated from DFT. The chemical potential differences depend on sample growth conditions, and their ranges are bounded by the formation enthalpy $\Delta H_f$ of Bi-227 (without defects) such as $\Delta H_f = 2\Delta\mu_{Bi} + 2\Delta\mu_{Ir} + 7\Delta\mu_O$, where $\Delta\mu_O$ is the chemical potential difference of two O atoms from an $O_2$ molecule in the gas phase. The DFT-calculated formation enthalpy of Bi-227 from the elemental Bi, Ir, and $O_2$ gas molecule is -15.08 eV. Therefore, the minimum values of $\Delta\mu_{Bi}$, $\Delta\mu_{Ir}$, and $\Delta\mu_O$ are -7.54, -7.54, and -2.15 eV, respectively, since their maximum values are zero. These DFT-calculated numbers assume $T = 0$. To include the temperature and oxygen pressure effect, we considered $\Delta\mu_O(T,P) = \Delta\mu_O(T,P_0) + (1/2)k_B T \ln(T/T_0)$, where the first term is the chemical potential difference of oxygen at $T$ and pressure $P_{O2} = 1$ atm which can be obtained by applying the ideal gas law and the standard tabulated values for the $O_2$ gas ($T_0 = 298$ K, $P_{O2} = 1$ atm).[74]

## Acknowledgements

We thank Professor J. Wang, Dr. Y. Wang, Prof. Y. Lan, Dr. C.-H. Chen and S. Cheema for fruitful discussions. S. X. Z. acknowledges start-up support from Indiana University (IU) College of Arts and Sciences. J. J. H. and Y. X. were supported for the transport measurements by the U.S. Department of Energy, Office of Basic Energy Sciences, Division of Materials Sciences and Engineering under award DOE DE-FG02-08ER46532. K.P. is grateful to John W. Villanova for creating the initial pyrochlore structure and to Denis Demchenko for an advice in computation of formation energy, and was supported by U.S. National Science Foundation DMR-1206354, San Diego Supercomputer Center (SDSC) Comet and Gordon under DMR060009N, and Advanced Research Computing at Virginia Tech. We thank the IU Nanoscale Characterization Facility for access to the scanning electron microscope. XPS instrument at Nanoscale Characterization Facility of IU Nanoscience Center was funded by NSF Award DMR MRI-1126394. The X-ray diffraction facilities at the IU Molecular Structure Center was supported by NSF Grant No. CHE-1048613. The work at Los Alamos National Laboratory was supported by the NNSA's Laboratory Directed Research and Development Program and was performed, in part, at the Center for Integrated Nanotechnologies, an Office of Science User Facility operated for the U.S. Department of Energy (DOE) Office of Science. Los Alamos National Laboratory, an affirmative action equal opportunity employer, is operated by Los Alamos National Security, LLC, for the National Nuclear Security Administration of the U.S. Department of Energy under contract DE-AC52-06NA25396.

**Author contributions statement**

W.C.Y., W.K.Z. and M.S. carried out thin film growth. W.C.Y. performed XRD and EDX measurements and XPS analysis. W.K.Z. carried out XRD characterizations. Y.T.X. conducted transport measurements and data analysis under the supervision of J.J.H. K.P. carried out DFT calculations and analysis. A.P.C. performed XRD characterizations under the supervision of Q.X.J. Y.L. conducted XPS measurements and guided W.C.Y. on XPS analysis. Z.L. carried out TEM characterizations and analysis under the supervision of N.L. L.H. and J.A.A. assisted on EDX characterizations. S.X.Z., J.J.H., W.K.Z., W.C.Y., and K.P. prepared the manuscript. All authors participated in discussion and reviewed the manuscript. S.X.Z. conceived and directed the overall project.

## Additional information

**Supplementary information** accompanies this paper at http://www.nature.com/srep

**Competing financial interests:** The authors declare no competing financial interests.

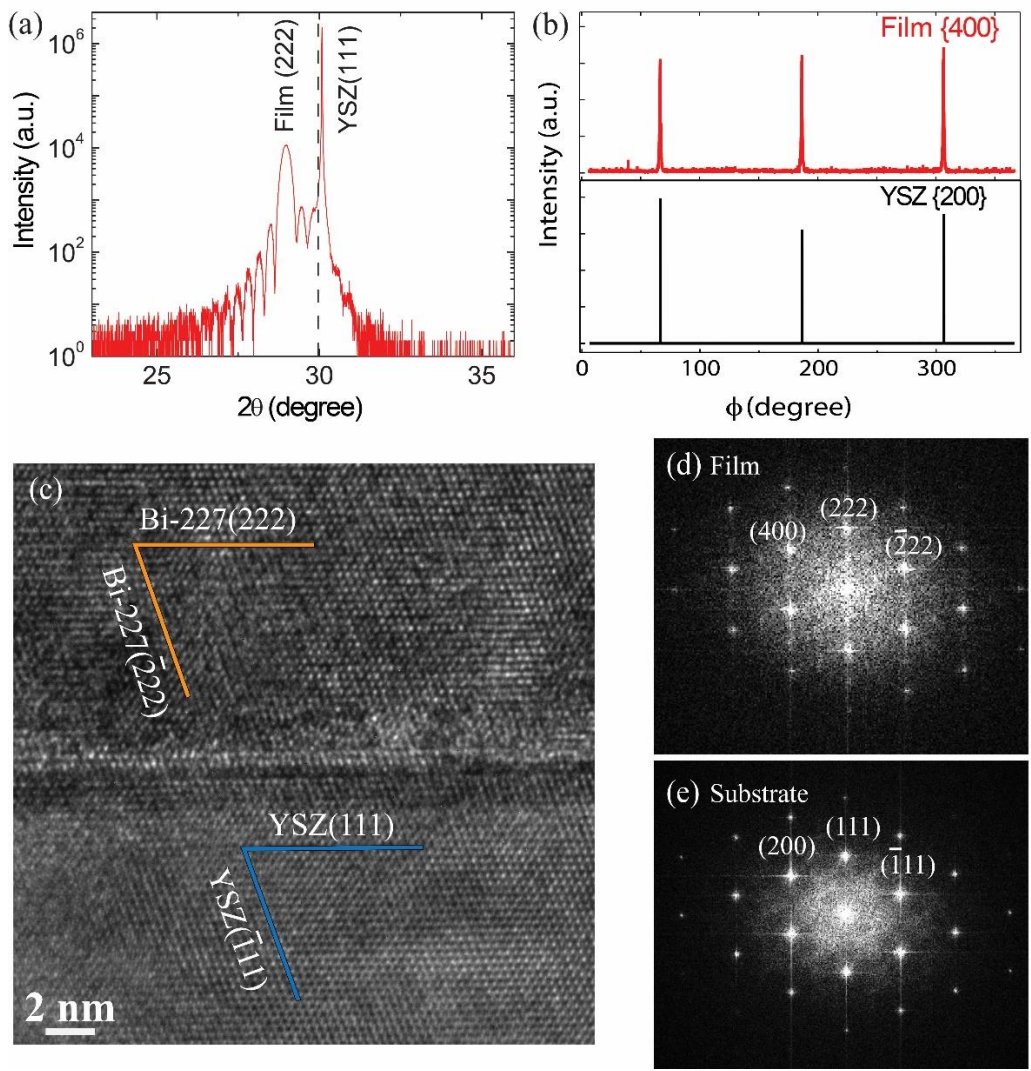

**Figure 1.** X-ray diffraction characterizations and TEM studies of a representative Bi-227 thin film grown at $P_{O2}$ =50 mTorr using Ir/Bi@1 target (f1): (a) *2θ-ω* scan around YSZ (111) reflection showing the (111) and (222) peaks of the substrate and thin film, respectively. The dashed line indicates the (222) peak position of the bulk single crystal Bi-227.[14,29] (b) in-plane *φ*-scan for the {200} of YSZ substrate and the {400} of thin film. (c) Cross-sectional TEM image taken at the interface of thin film and substrate; Fast Fourier Transform images from (d) thin film and (e) substrate.

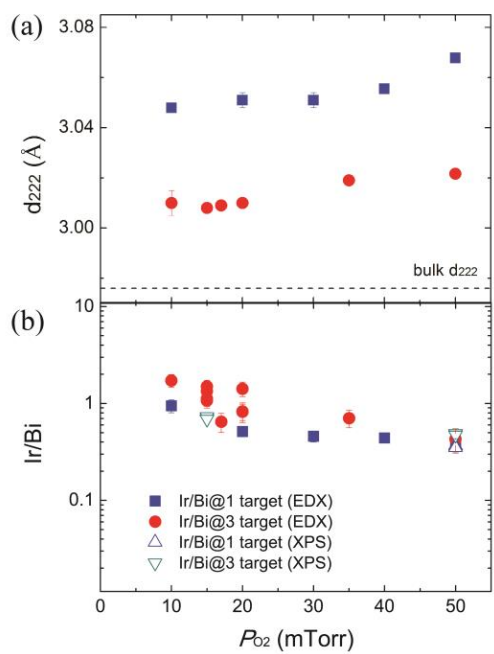

**Figure 2.** (a) lattice spacing $d_{222}$ versus oxygen pressure $P_{O2}$. The error bars were estimated according to the diffraction peaks. (b) Ir/Bi ratio versus $P_{O2}$ (multiple data points at the same pressure were taken at different spots/areas of the sample). The error bars were determined based on the EDX analysis in the AZtech software.

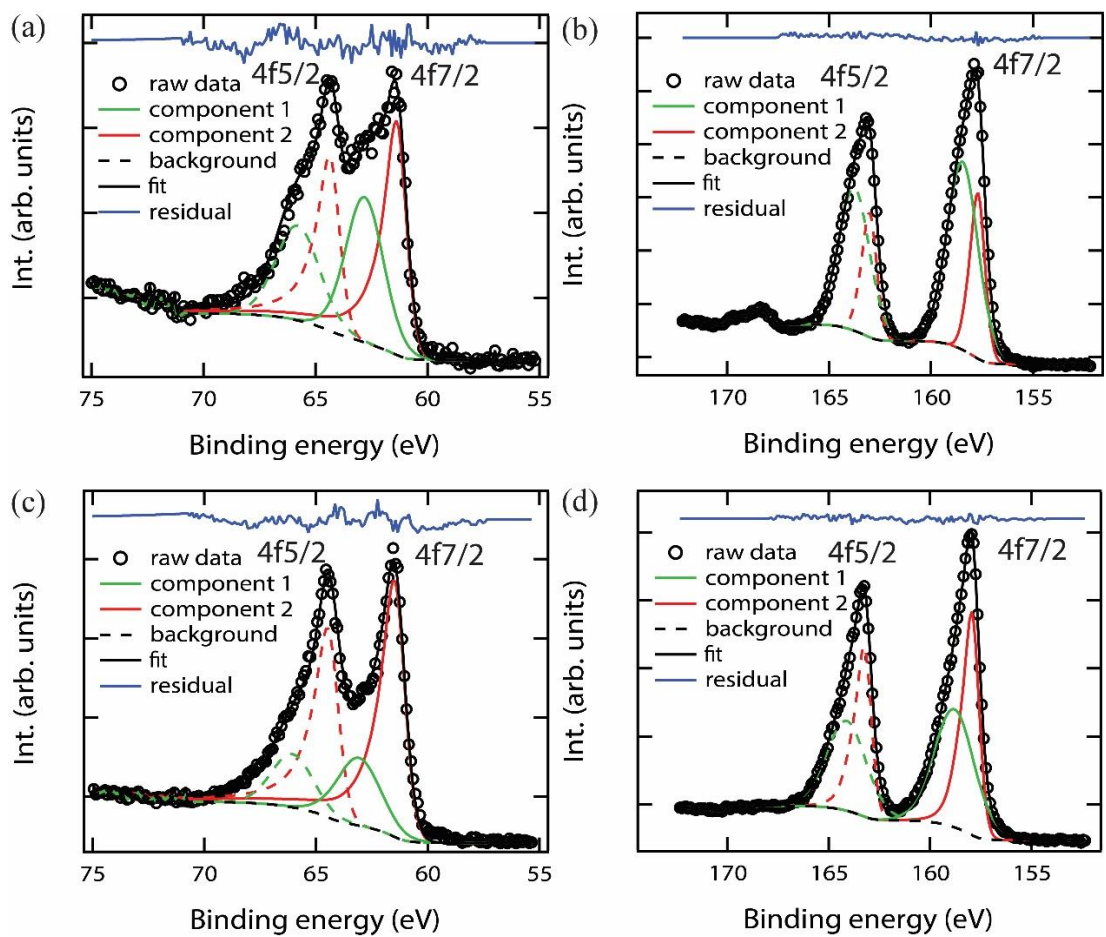

**Figure 3.** XPS (a) Ir $4f_{5/2}$ and $4f_{7/2}$ spectra and (b) Bi $4f_{5/2}$ and $4f_{7/2}$ spectra taken on film f1. (c) Ir $4f_{5/2}$ and $4f_{7/2}$ spectra and (d) Bi $4f_{5/2}$ and $4f_{7/2}$ spectra taken on film f3. The spectra were fitted using two components: a symmetric component (higher binding energy peak) described by the Gaussian-Lorentz profile GL(m), and an asymmetric component (lower binding energy peak) by the convolution of Gelius profile A(a,b,n) and Gaussian-Lorentz profile GL(m). The parameters a and b are optimized as described in the supplementary information.

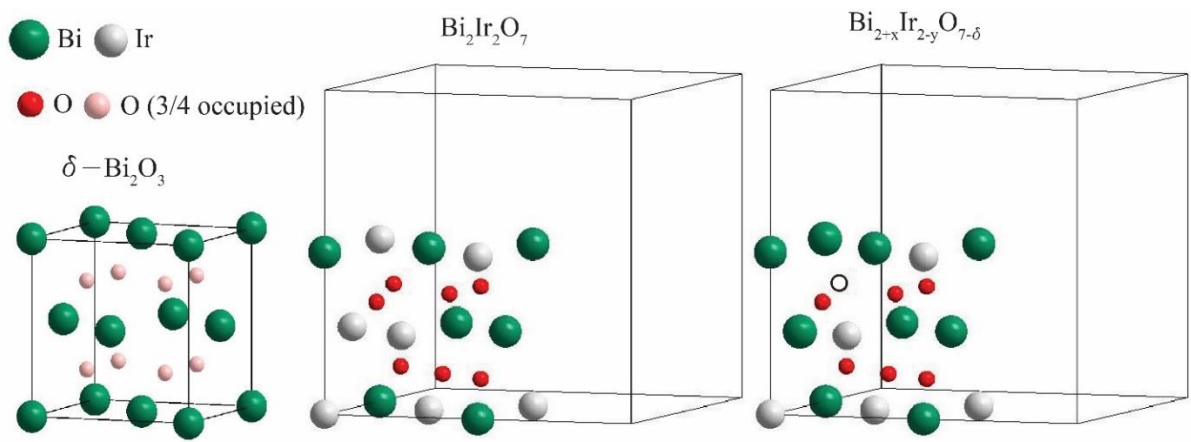

**Figure 4**. A unit cell of cubic δ–$Bi_2O_3$, 1/8 unit cell of $Bi_2Ir_2O_7$ and $Bi_{2+x}Ir_{2-y}O_{7-\delta}$. In the $Bi_{2+x}Ir_{2-y}O_{7-\delta}$ shown above (as an example), two Ir atoms are replaced by Bi atoms and one oxygen atom is missing (represented by the open circle).

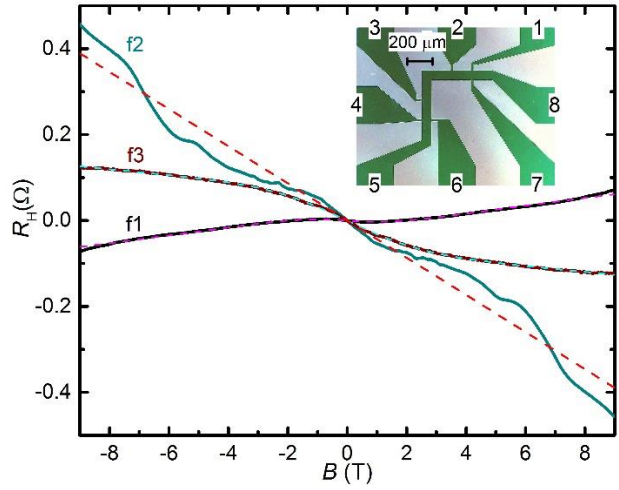

**Figure 5.** Hall resistance $R_H$ vs $B$ at $T = 0.39$ K for films f1, f2 and f3 (solid lines, films as indicated), with fits to a two-carrier model (f1, f3, dashed lines) or an averaging single-carrier model (f2, dashed line). The polarities are chosen such that a negative slope corresponds to positive charge carriers. The inset represents the L-shaped Hall bar geometry used on all 3 films (contacts 5 and 8 are current contacts, contacts 1, 2, 3, 4, 6, 7 are voltage contacts for measurement of $R_H$ and $R_\square$).

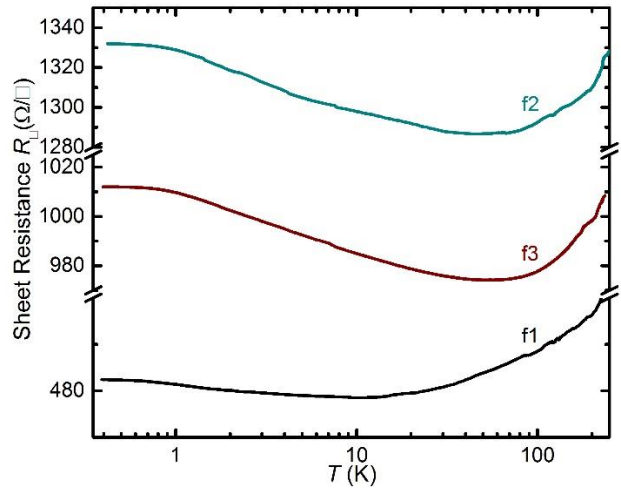

**Figure 6.** Sheet resistance $R_\square$ vs $T$ at $B=0$ for films f1, f2 and f3 (films as indicated).

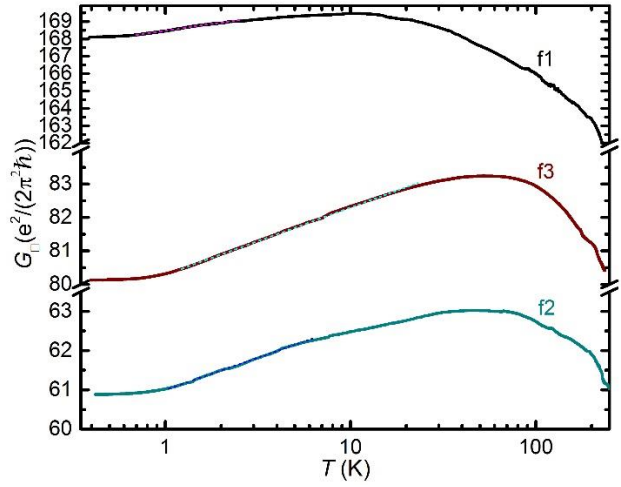

**Figure 7.** Sheet conductance $G_\square$ vs $T$ at $B=0$ for films f1, f2 and f3 (solid lines, films as indicated), with fits to the low-$T$ data using Equation (1) (dashed lines).

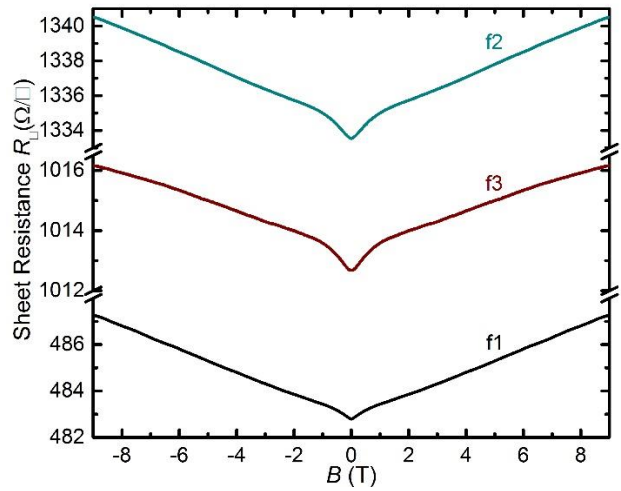

**Figure 8.** Sheet resistance $R_\square$ vs $B$ at $T = 0.39$ K for films f1, f2 and f3 (films as indicated). A positive MR attributed to antilocalization is visible for $B < 1$ T, and a linear positive MR appears for $B > 1.5$ T.

| Bi: Ir antisite [Bi interstit.] | Case 1 | Case 2 | Case 3 | Case 4 | Case 5 |
|---|---|---|---|---|---|
| $\Delta\mu_{Bi}^{T=0}$ (eV) | 0 | -3.77 | 0 | -2.20 | 0 |
| $\Delta\mu_{Ir}^{T=0}$ (eV) | -7.54 | -3.77 | -5.79 | -2.20 | 0 |
| $\Delta\mu_{O}^{T=0}$ (eV) | 0 | 0 | -0.5 | -0.9 | -2.15 |
| Form. Eng. $\Delta E_f$ (eV) | -7.47 [4.86] | -4.04 [6.50] | -5.72 [4.86] | -4.04 [4.93] | 4.04 [4.86] |
| Notes | $\Delta E_f$ at $T$=550 C, $P$=50mTorr | $\Delta E_f$ at $T$=550 C, $P$=15mTorr | $\Delta E_f$ at $T$=550 C, $P$=50mTorr | $\Delta E_f$ at $T$=550 C, $P$=15mTorr | $\Delta E_f$ at $T$=0 |

**Table 1.** Calculated formation energies $\Delta E_f$ of a single Bi$_{Ir}$ defect and a Bi$_{int}$ in the pyrochlore structure of Bi-227 with the chemical potential differences of Bi, Ir, and O, $\Delta\mu_{Bi}^{T=0}$, $\Delta\mu_{Ir}^{T=0}$, and $\Delta\mu_{O}^{T=0}$ at $T$=0 in five different cases. The un-bracketed (bracketed) numbers are the formation energies of the Bi$_{Ir}$ (Bi$_{int}$) defect. Here the minimum value of both $\Delta\mu_{Bi}^{T=0}$ and $\Delta\mu_{Ir}^{T=0}$ is -7.54 eV, and the minimum value of $\Delta\mu_{O}^{T=0}$ is -2.15 eV, considering the zero-temperature DFT-calculated Bi-227 formation enthalpy per unit (consisting of 2 Bi, 2 Ir, and 7 O atoms) to be -15.08 eV. For the calculation of $\Delta E_f$, the temperature and oxygen pressure dependence of $\Delta\mu_O$ was included except for Case 5.

Supplementary Information

**Epitaxial thin films of pyrochlore iridate Bi$_{2+x}$Ir$_{2-y}$O$_{7-\delta}$: structure, defects and transport properties**

W. C. Yang[1,+], Y. T. Xie[2,+], W. K. Zhu[1,+], K. Park[2], A. P. Chen[3], Y. Losovyj[4], Z. Li[1,3], H. Liu[1], M. Starr[1], J. A. Acosta[1], C. G. Tao[2], N. Li[3], Q. X. Jia[3,5], J. J. Heremans[2], and S. X. Zhang[1,*]

[1]Department of Physics, Indiana University, Bloomington, Indiana 47405, USA
[2]Department of Physics, Virginia Tech, Blacksburg, Virginia 24061, USA
[3]Center for Integrated Nanotechnologies, Los Alamos National Laboratory, Los Alamos, 87545 USA
[4]Department of Chemistry, Indiana University, Bloomington, Indiana 47405, USA
[5]Department of Materials Design and Innovation, University at Buffalo, The State University of New York, Buffalo, NY 14260, USA
* sxzhang@indiana.edu
[+] these authors contributed equally to this work

1. **X-ray diffraction (XRD) characterization of thin films**

Figure S1 shows the XRD 2θ-ω patterns of the YSZ (111) substrate and some representative thin films. The (222), (333) and (444) peaks of the Bi-227 films are identified. In particular, the detection of (333) peak suggests good film quality. The Ir (111) peak is observed in the films grown at 10mTorr and 1mTorr, indicating the formation of iridium metal impurity at low $P_{O2}$. The film at 1mTorr is mainly composed of Ir metal, consistent with the EDX measurement (Figure S2(c)). The peaks denoted by '*' are due to Cu k$_\beta$ and the other tiny sharp peaks are either due to the instrument or the substrate as they are observed in the bare substrate as well. Since the (222) peaks of some films are too close to the substrate (111) peak, preventing from an accurate determination of the diffraction angles, the $d_{222}$ values in Figure 2(a) were calculated based on the (444) peaks, i.e. $d_{222} = 2d_{444}$.

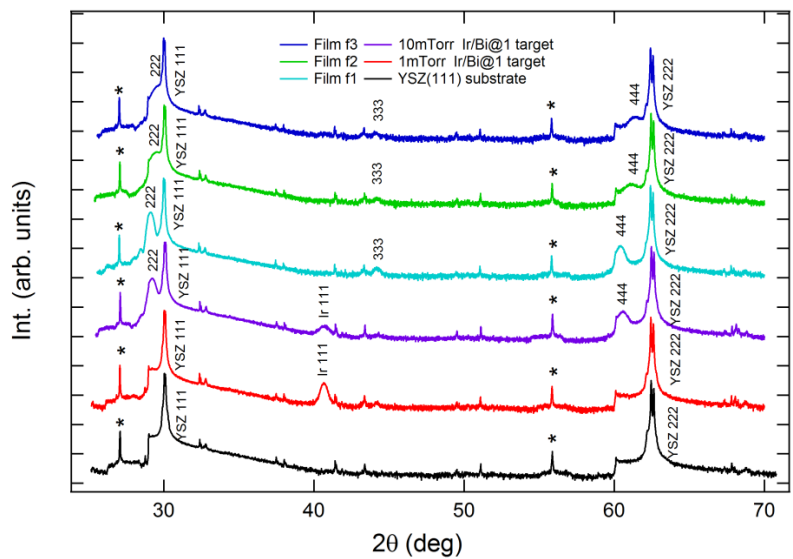

*Figure S1.* XRD 2θ–ω scans of the YSZ substrate and the films grown at: $P_{O2}$= 50 mTorr, Ir/Bi@1 target (f1); $P_{O2}$ = 50 mTorr, Ir/Bi@3 target (f2); $P_{O2}$ = 15 mTorr, Ir/Bi@3 target (f3); $P_{O2}$= 10 mTorr, Ir/Bi@1 target; and $P_{O2}$= 1 mTorr, Ir/Bi@1 target. The vertical axis is in logarithmic scale and all patterns are corrected/calibrated against the substrate peak with a lattice constant of $a_{ysz}$=5.125 Å to determine the d-spacings of thin films.

## 2. Energy-dispersive X-ray spectroscopy study of thin films

Energy-dispersive X-ray spectroscopy (EDX) was employed to characterize the chemical composition of thin films. Figure S2(a) shows a scanning electron microscope (SEM) image of a representative thin film (10 mTorr with an Ir/Bi@3 target). Figure S2(b) is the EDX spectrum which confirms the existence of Ir and Bi in the film. The Y, Zr and Hf signals are from the YSZ substrate which often contains Hf impurity. The atomic ratio of Ir/Bi is calculated in the AZtech software. Figure S2(c) is the EDX spectrum of a film grown at $P_{O2}$ = 1 mTorr which has a negligible Bi signal.

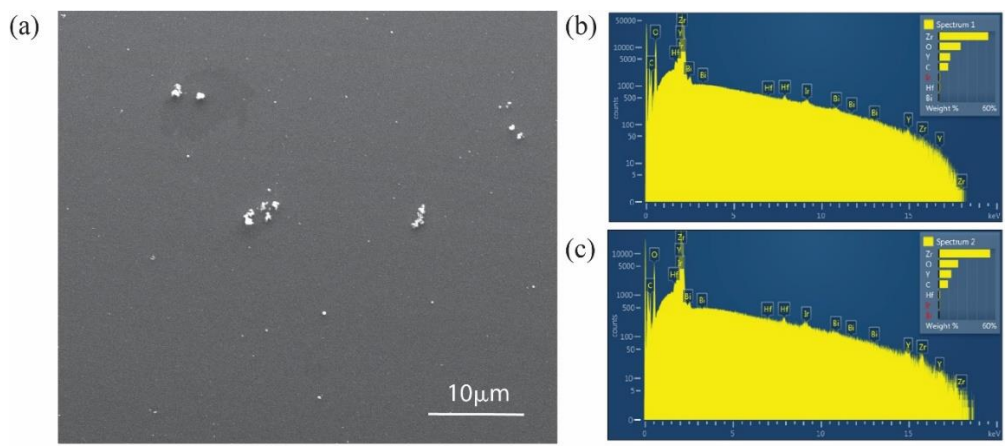

*Figure S2.* (a) An SEM image and (b) EDX spectrum of a thin film grown at $P_{O2} = 10$ mTorr using an Ir/Bi@3 target. (c) EDX spectrum of a film grown at $P_{O2} = 1$ mTorr using an Ir/Bi@1 target.

## 3. X-ray photoelectron spectroscopy (XPS) characterization of film f2 ($P_{O2}$ = 50 mTorr, Ir/Bi@3 target)

Figure S3 (a) and (b) show the XPS Ir and Bi spectra taken on film f2, respectively. The spectra were fitted using the procedure described in section 5.

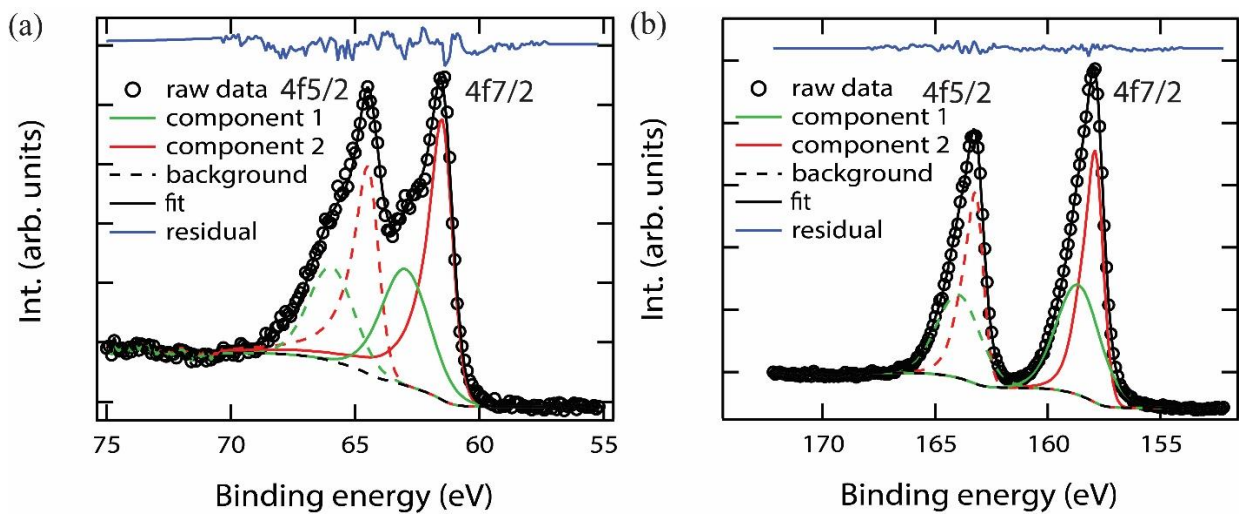

*Figure S3.* XPS (a) Ir $4f_{5/2}$ and $4f_{7/2}$ spectra and (b) Bi $4f_{5/2}$ and $4f_{7/2}$ spectra of film f2. The spectra are fitted using a Gaussian lineshape GL(m) and a convolution of Gelius profile A(a,b,n) and GL(m).

## 4. Comparison of XPS spectra between films and bulk $IrO_2$, $Bi_2O_3$ powder

XPS measurements were taken on standard $IrO_2$ and $Bi_2O_3$ powder as references. The peaks from thin films are on the higher side of the spectra comparing to $IrO_2$ and $Bi_2O_3$, suggesting lower oxidation states in the thin film samples.

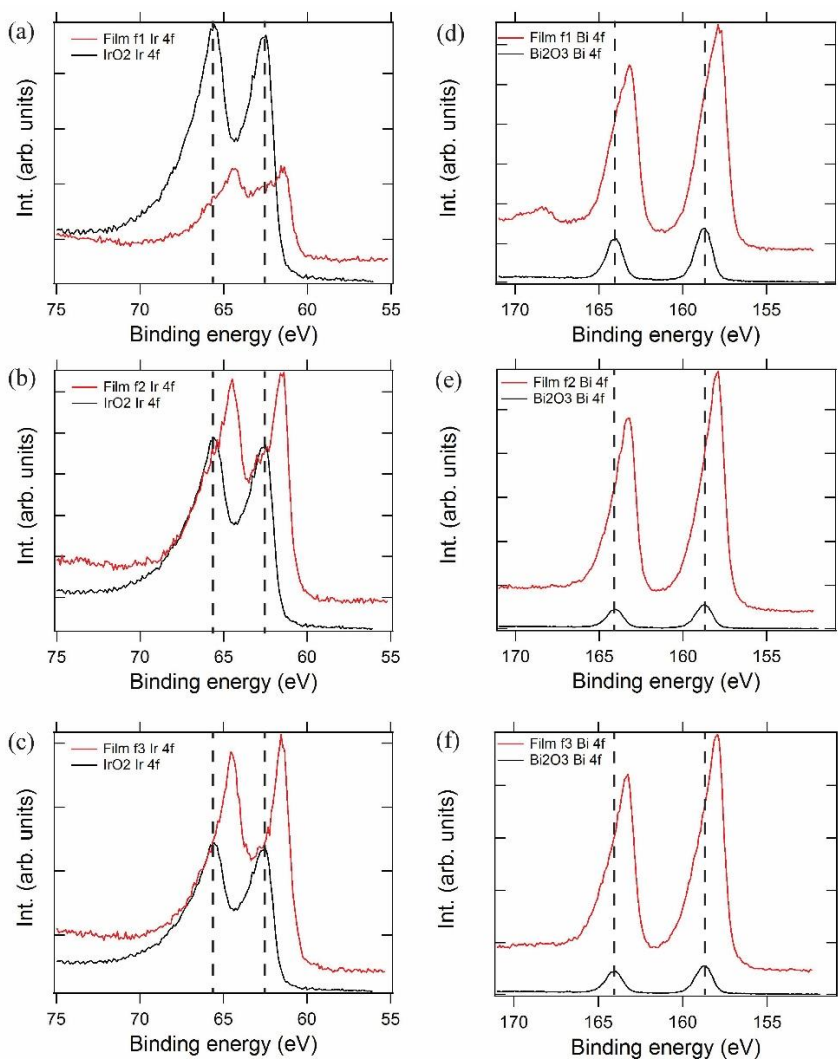

*Figure S4.* Comparison of XPS spectra of $IrO_2$ and $Bi_2O_3$ with compound (a) $IrO_2$ and film f1 (b) $IrO_2$ and film f2 (c) $IrO_2$ and film f3 (d) $Bi_2O_3$ and film f1 (e) $Bi_2O_3$ and film f2 (f) $Bi_2O_3$ and film f3. The dashed lines indicate the binding energies of 4f5/2 and 4f7/2 states.

## 5. Fitting of XPS spectra

The XPS spectra were fitted using CasaXPS software. The high binding energy peak was fitted using a symmetric Gaussian lineshape GL(m) while the lower one was fitted by an asymmetric lineshape. The asymmetric lineshape was described by the convolution of Gelius profile A(a,b,n) and Gaussian GL(m), in which the asymmetric part is characterized by A(a,b,n). In a typical fitting process, we choose one combination of parameters a and b in the Gelius profile and run the fitting from which we obtain the residual spectra, i.e. the difference between the experimental data and the fitting spectra. By repeating the procedure with different combinations of a and b, a residual STD as a function of parameter a and b is obtained. The parameters a and b are manually tuned in order to yield a minimum value in the residual STD curve. Figure S5 gives the residual STD versus parameter b with different a for f1 Ir spectrum, the best fit parameters are obtained when the residual function reaches a minimum value, in this case a=0.4 and b=0.6.

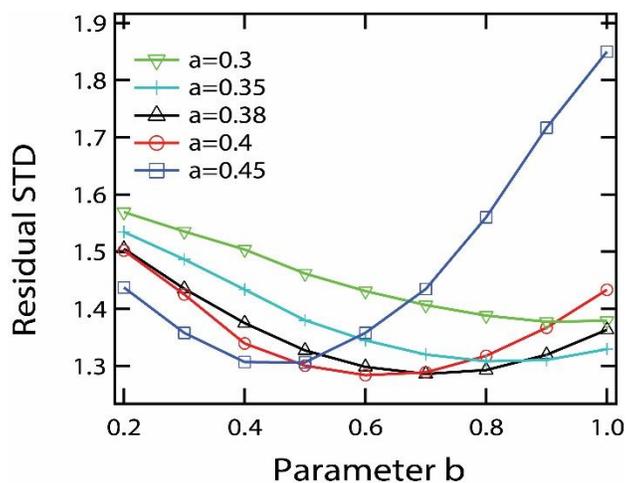

*Figure S5. Residual STD versus parameter b with different parameter a for the Ir spectrum of film f1.*

## 6. Unit cells of $Bi_2Ir_2O_7$ with and without point defect

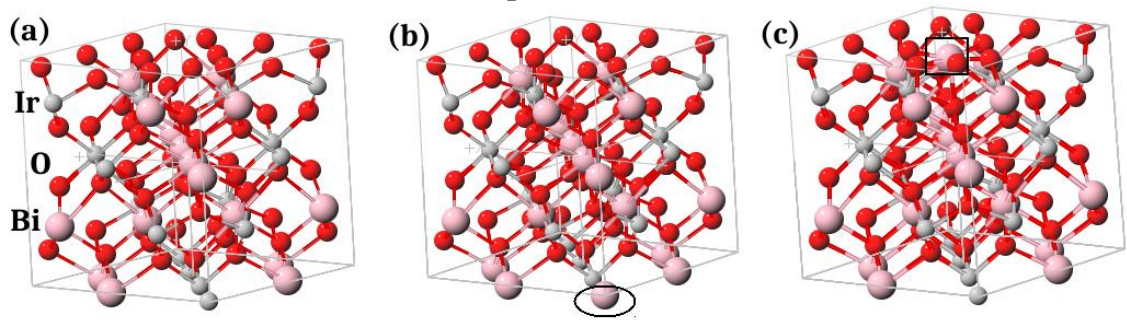

*Figure S6.* The 88-atom unit cells of (a) a perfect crystal, (b) one $Bi_{Ir}$ antisite (indicated by the circled Bi atom at bottom), and (c) one interstitial Bi atom among the O atoms (indicated by the boxed Bi atom on top). Although a unit cell of perfect crystal consists of 22 atoms, for clear comparison with (b) and (c), an 88-atom unit cell is shown for (a).